\documentclass[10pt,aps,twocolumn,showpacs,pra]{revtex4} 
\usepackage{amssymb,amsfonts,amsmath,wrapfig,mathrsfs,mathbbol} 
\usepackage{graphicx}

\begin{document}
\title{Manipulation of single-photon states encoded in transverse spatial modes:\\ possible and impossible tasks}
\author{Gabriel F. Calvo and Antonio Pic\'{o}n}
\affiliation{
Grup de F\'{\i}sica Te\`{o}rica, Universitat Aut\`{o}noma de Barcelona, 08193 Bellaterra (Barcelona), Spain
}
\date{\today}
\begin{abstract}
Controlled generation and manipulation of photon states encoded in their spatial degrees of freedom is a crucial ingredient in many quantum information tasks exploiting higher-than-two dimensional encoding. Here, we prove the impossibility to arbitrarily modify $d$-level state superpositions (qu$d$its) for $d>2$, encoded in the transverse modes of light,  with optical components associated to the group of symplectic transforms (Gaussian operations). Surprisingly, we also provide an explicit construction of how non-Gaussian operations acting on mode subspaces do enable to overcome the limit $d=2$. In addition, this set of operations realizes the full SU(3) algebra.
\end{abstract}

\maketitle 
\section{INTRODUCTION}
\label{sect:intro}
\vspace*{-0.2cm}
Most promising approaches for scalable quantum communication (QC) rely on the use of photons as the main carriers of information among remote nodes of quantum networks, where matter-based quantum memories are located~\cite{Duan01,Sherson,Rosenfeld,Collins,Boozer}. Photons, besides being the natural candidate for QC due to their long decoherence time and the relative ease with which they can be manipulated, can actually encode multiple quantum bits of information (qubits) into various degrees of freedom. These include frequency, polarization, linear momentum and orbital angular momentum~\cite{OAMbook}. The possibility of simultaneously exploiting these degrees of freedom~\cite{Barreiro05,Walborn06} is becoming increasingly appealing for the faithful mapping of quantum states between light and matter~\cite{Rosenfeld}. A fundamental question then arises: {\em What are the most general photon state manipulations allowed by benchmark optical components?} It is of paramount importance to obtain a clear representation of all such state mappings to further develop a truly multi-degree-of-freedom photon state engineering.
\par
Here, we address the problem of whether, by resorting to the symplectic group of optical transformations  on spatial transverse-field modes, it is possible to perform arbitrary manipulations on photon states encoded in large, but finite, $d$-dimensional superpositions of these modes (qu$d$its). This is relevant for non-dichotomic QC protocols, which include those exploiting multimode squeezing~\cite{Lassen} and the orbital angular momentum (OAM) of light~\cite{Mair,Leach02,Vaziri03,Molina04,Langford,Gibson,Kumar, Stutz}. For OAM, one of its main distinguishing features is the access to, in principle, an infinite-dimensional Hilbert space expanded by cylindrically-symmetric paraxial eigenmodes (e.g. the Laguerre-Gaussian basis)~\cite{Calvo05,Aiello05,Calvo06,Calvo07}. Spatial encoding conveys several independent channels of information that could be very useful in quantum cryptographic schemes with larger alphabets~\cite{Qutrit} and security enhancement against eavesdropping~\cite{Security}. Even for quantum computation applications, the high-dimensional aspect would enable to optimize certain computing arquitectures~\cite{Greentree}. 
\par
A necessary condition to perform arbitrary unitary operations on a pure quantum state $\vert\psi\rangle=  \sum_{j=1}^{d}\alpha_{j}\vert j\rangle$, consisting of a $d$-dimensional superposition of orthogonal eigenmodes $\vert j\rangle$, is to modify in a controlled way each of the complex coefficients $\alpha_{j}$. In most of the experimental realizations oriented towards the use of spatial degrees of photons for high-dimensional encoding, phase holograms and reconfigurable spatial light modulators have been employed to approximately manipulate specific combinations of optical transverse modes~\cite{OAMbook}. In practice, however, these elements do not strictly preserve paraxiality but, rather, behave as non-unitary transformations, thus constituting a source of mode noise that eventually destroys the desired large, but finite, multidimensionality of the quantum states to be exploited. 
Our first main result shows that when these, or any combination of, optical elements belong to the group of symplectic transformations (which include Gaussian operations), it is impossible to arbitrarily modify
single-photon qu$d$it states for $d>2$, via unitary operations generated by those transforms. Hence, a clear motivation emerges: {\em Is it possible to find transformations on paraxial modes which allow one to really overcome the limit $d=2$?} Our second main result provides a positive answer to this question; we present a set of non-Gaussian operations that truly enable us to arbitrarily manipulate (up to global phases) single-photon qutrit ($d=3$) states. Furthermore, this set of operations constitutes a SU(3) algebra.
\par 
The paper is organized as follows: Section \ref{sec:SF} gives a brief summary of the formalism on symplectic groups and transformations in the optical phase space. In Section \ref{sec:MOPM} we introduce and characterize the most general representation of unitary (metaplectic) operators corresponding to all possible optical symplectic transformations that can be performed on transverse field modes. Section \ref{sec:SQGMO} provides the first main result of our paper; we prove that, via the group of symplectic transforms acting on superpositions of paraxial modes, it is impossible to implement operations that change arbitrary  qudit states onto any other qudit for $d>2$. In Section \ref{sec:NGGates} we extend our analysis to non-Gaussian operations on these modes and present new routes towards the aim of truly manipulating arbitrary single-photon qu$d$its. Section \ref{sec:CONCLU} concludes the paper with a discussion of alternative approaches to implement controlled gates on single-photons using more than one of their degrees of freedom. A simple optical scheme for a CNOT gate exploiting OAM and polarization, is proposed.
\par
\section{SYMPLECTIC GROUP FORMALISM}
\label{sec:SF}
\vspace*{-0.2cm}
To put in context the class of optical transformations referred to above, it is necessary to start by introducing the symplectic formalism that will be used extensively throughout the paper. We first recall that the dynamics of classical and quantum Hamiltonian systems has an underlying symplectic structure. Symplectic methods have been applied in the theory of elementary particles, condensed matter, accelerator and plasma physics, oceanographic and atmospheric sciences and in optics~\cite{Kim83,Guillemin,Torre}. Fundamental to all of them is the phase space picture. Any classical system with $n$ degrees of freedom is described by a set of pairs $q_{j},p_{j}$ ($j=1,2,\ldots,n$) of mutually conjugate canonical variables. In the quantum domain one can associate to these variables the irreducible set of canonical Hermitian operators $\hat{q}_{j},\hat{p}_{j}$. The basic kinematic structure is provided by Poisson brackets in the former case and by the Heisenberg commutation relations in the latter. By assembling the canonical variables and operators into $2n$-component vectors ${\boldsymbol \xi}=(q_{1},q_{2},\ldots,q_{n},p_{1},p_{2},\ldots,p_{n})$ and $\hat{\boldsymbol\xi}=(\hat{q}_{1},\hat{q}_{2},\ldots,\hat{q}_{n},\hat{p}_{1},\hat{p}_{2},\ldots,\hat{p}_{n})$, the Poisson brackets and the Heisenberg commutation relations can be cast, respectively, as $\{\xi_{\alpha},\xi_{\beta}\}=\Omega_{\alpha,\beta}$ and $[\hat{\xi}_{\alpha},\hat{\xi}_{\beta}]=i\Omega_{\alpha,\beta}$, ($\alpha,\beta=1,2,\ldots,2n$), where
\begin{eqnarray}
\Omega =\left( \begin{array}{cc}
\mathbf{0}_{n\times n} & \mathbf{1}_{n\times n} \\
-\mathbf{1}_{n\times n} & \mathbf{0}_{n\times n}
\end{array} \right) ,
\label{eq:SymplecticMatrix}
\end{eqnarray}
is the $2n$-dimensional symplectic metric matrix. Of particular relevance are the real linear canonical transformations among quantum (classical) canonical quantities~\cite{Arvind95a}. They preserve the Heisenberg (Poisson) relations and are represented by symplectic matrices $S:\hat{\boldsymbol \xi} \rightarrow \hat{\boldsymbol \xi}' = S \hat{\boldsymbol \xi}$, obeying the condition $S \Omega S^{T} = \Omega$. The set of all such $2n$-dimensional real matrices forms the $(2n^{2}+n)$-parameter non-compact symplectic group Sp($2n,\mathbf{R}$). 
\par
The power of symplectic formalism becomes apparent in the following general setting. Let $\mathcal{H}$ denote the Hilbert space of $n$-mode states $\hat{\rho}$ on which $\hat{\xi}_{\alpha}$ act. Given that for any $S\in\textrm{Sp}(2n,\mathbf{R})$ the Hermiticity properties and commutation relations of the $\hat{\xi}_{\alpha}$ are conserved, and since $\hat{\xi}_{\alpha}$ act irreducibly on $\mathcal{H}$, it follows from Stone-von Neumann theorem that one can define unitary operators $\hat{U}(S)$ on $\mathcal{H}$, implementing $S \Omega S^{T} = \Omega$, such that~\cite{Arvind95a}
\begin{eqnarray}
\hat{U}^{\dagger}(S) \hat{\boldsymbol \xi}\hat{U}(S) = S \hat{\boldsymbol \xi}\,. 
\label{eq:StonevonNeumannTh}
\end{eqnarray}
Clearly, all possible transformations $S$ are mapped, up to a sign ambiguity, onto $\hat{U}(S)$. Hence, all density operators $\hat{\rho}\in\mathcal{H}$ transform under $\hat{U}(S)$ as $\hat{\rho}'=\hat{U}(S)\hat{\rho}\,\hat{U}^{\dagger}(S)$. On the level of Wigner functions $W({\boldsymbol \xi})$, this transformation acquires a strikingly simple form: $W'({\boldsymbol \xi})=W(S^{-1}{\boldsymbol \xi})$.
\par
\section{METAPLECTIC OPERATIONS ON PARAXIAL MODES}
\label{sec:MOPM}
Within classical and quantum optics, the symplectic formalism has been extensively used both in studying mode-mapping properties of lossless first-order (paraxial or {\bf ABCD}) systems~\cite{Simon00a,Simon00b,Calvo05,Calvo06,Alieva} and in characterization of continuous-variable entanglement~\cite{Gaussian}. An important class of symplectic transforms is that of two-mode systems represented by the symplectic group $\textrm{Sp}(4,\mathbf{R})$. For instance, bipartite Gaussian operations, which preserve the Gaussian character of the Wigner functions, belong to $\textrm{Sp}(4,\mathbf{R})$. Let us make explicit the form of all possible $\hat{U}(S)$ when $S\in\textrm{Sp}(4,\mathbf{R})$. They give rise to the unitary metaplectic representation of Sp(4,$\mathbf{R}$) acting on $\mathcal{H}$. All these unitary operations  $\hat{U}(S)$ are generated by ten Hermitian operators $\hat{\boldsymbol{\mathcal{J}}}$, quadratic in $\hat{\boldsymbol \xi}$, that can be split into two sets~\cite{Arvind95b}: passive and active generators. The passive set encompasses the maximal compact subgroup U$(2)$: 
\begin{eqnarray}
\hat{\mathcal{L}}_{o}&=& \frac{1}{2} \left(\hat{a}_{x}^{\dagger}\hat{a}_{x}+\hat{a}_{y}^{\dagger}\hat{a}_{y}\right) , \nonumber \\
\hat{\mathcal{L}}_{x}&=& \frac{1}{2} \left(\hat{a}_{x}^{\dagger}\hat{a}_{x}-\hat{a}_{y}^{\dagger}\hat{a}_{y}\right) , \nonumber \\
\hat{\mathcal{L}}_{y}&=& \frac{1}{2} \left(\hat{a}_{x}^{\dagger}\hat{a}_{y}+\hat{a}_{y}^{\dagger}\hat{a}_{x}\right) , \nonumber \\
\hat{\mathcal{L}}_{z}&=& - \frac{i}{2} \left(\hat{a}_{x}^{\dagger}\hat{a}_{y}-\hat{a}_{y}^{\dagger}\hat{a}_{x}\right) .
\label{eq:passiveoperators}
\end{eqnarray}
Here, $\hat{a}_{j}= (\hat{q}_{j}+i\hat{p}_{j})/\sqrt{2}$, (respectively $\hat{a}_{j}^{\dagger}$), $j=x,y$, are the two annihilation (creation) operators for all orthogonal transverse modes.  The passive operators~(\ref{eq:passiveoperators}) have the form of the well-known Stokes operators. They obey the usual commutation relations $[\hat{\mathcal{L}}_{i},\hat{\mathcal{L}}_{j}]= i\,\varepsilon_{ijk}\hat{\mathcal{L}}_{k} $ ($i,j,k=x,y,z$), with $\hat{\mathcal{L}}_{o}$ being the only commuting element in U(2).  
\par
The active set is responsible for the noncompactness of Sp(4,$\mathbf{R}$):
\begin{eqnarray}
\hat{\mathcal{K}}_{x} &=& -\frac{1}{2} (\hat{a}^{\dagger}_{x}\hat{a}^{\dagger}_{y}+\hat{a}_{x}\hat{a}_{y})\, ,  \nonumber \\
\hat{\mathcal{K}}_{y} &=& \frac{1}{4} (\hat{a}^{\dagger \, 2}_{x} + \hat{a}^{2}_{x} - \hat{a}^{\dagger \, 2}_{y} - \hat{a}^{2}_{y})\, , \nonumber \\
\hat{\mathcal{K}}_{z} &=&  -\frac{i}{4} (\hat{a}^{\dagger \, 2}_{x} - \hat{a}^{2}_{x} + \hat{a}^{\dagger \, 2}_{y} - \hat{a}^{2}_{y})\, ,\nonumber \\
\hat{\mathcal{M}}_{x} &=& -\frac{i}{2} (\hat{a}^{\dagger}_{x}\hat{a}^{\dagger}_{y}-\hat{a}_{x}\hat{a}_{y})\, , \nonumber \\
\hat{\mathcal{M}}_{y} &=& \frac{i}{4} (\hat{a}^{\dagger \, 2}_{x} - \hat{a}^{2}_{x} - \hat{a}^{\dagger \, 2}_{y} + \hat{a}^{2}_{y})\, , \nonumber \\
\hat{\mathcal{M}}_{z} &=&  \frac{1}{4} (\hat{a}^{\dagger \, 2}_{x} + \hat{a}^{2}_{x} + \hat{a}^{\dagger \, 2}_{y} + \hat{a}^{2}_{y})\, ,
\label{eq:activeoperators}
\end{eqnarray}
satisfying the following commutators:
\begin{eqnarray}
[\hat{\mathcal{L}}_{i},\hat{\mathcal{K}}_{j}]= i \,\varepsilon_{ijk} \hat{\mathcal{K}}_{k} \, , \quad [\hat{\mathcal{L}}_{i},\hat{\mathcal{M}}_{j}] = i \,\varepsilon_{ijk} \hat{\mathcal{M}}_{k} \, ,\nonumber \\\nonumber [\hat{\mathcal{L}}_{o}, \hat{\mathcal{K}}_{j} \pm i \hat{\mathcal{M}}_{j}] = \mp (\hat{\mathcal{K}}_{j} \pm i \hat{\mathcal{M}}_{j}) \, , \quad  [\hat{\mathcal{K}}_{i},\hat{\mathcal{M}}_{j}] = i\,\delta_{ij} \hat{\mathcal{L}}_{o}\, , \\ \nonumber [\hat{\mathcal{K}}_{i},\hat{\mathcal{K}}_{j}] = [\hat{\mathcal{M}}_{i},\hat{\mathcal{M}}_{j}] = - i \, \varepsilon_{ijk} \hat{\mathcal{L}}_{k}\,  .
\end{eqnarray}
Below we elucidate the action of each of the ten generators on a spatial mode carrying OAM.
\par
As any arbitrary sequence of symplectic transformations $S_{m}$ is again another symplectic transformation $S=\prod_{m}S_{m}$, one concludes that the most general metaplectic operator $\hat{U}(S)$ corresponding to $S$ is represented by a single exponential of $i$ times real linear combinations of any of the above generators: 
\begin{eqnarray}
\hat{U}(S)=\exp\left(-i{\bf s}\cdot\hat{\boldsymbol{\mathcal{J}}}\right) ,
\label{eq:metaplectic}
\end{eqnarray}
with $\hat{\boldsymbol{\mathcal{J}}}\in\{\hat{\mathcal{L}},\hat{\mathcal{K}},\hat{\mathcal{M}}\}$, and ${\bf s}$ a ten-parameter vector. 
\par
\begin{figure}
\begin{center}
\hspace*{-0.0cm}
\includegraphics[width=80mm]{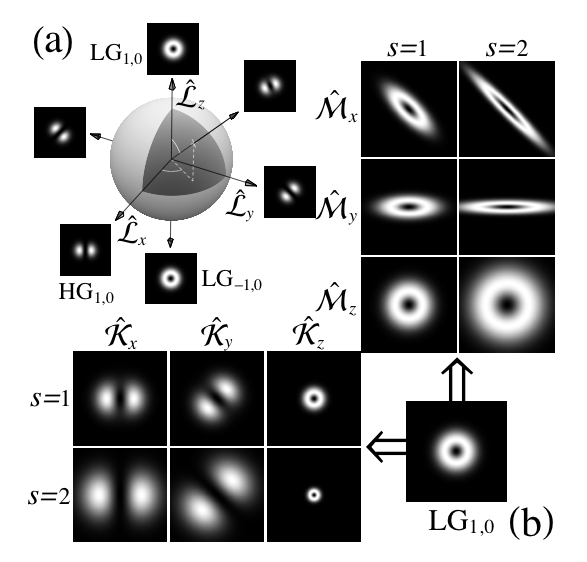}
\end{center}
\vspace*{-0.4cm}
\caption{\small Action of passive (a), and active (b) generators on a Laguerre-Gaussian mode $\textrm{LG}_{\ell=1,p=0}$. Profiles in (a) are mapped onto the first-order orbital Poincar\'{e} sphere. Profiles in (b) are given for increasing squeezing parameter $s$.}
\label{fig:GeneratorAction}
\end{figure}

When applied to photon number states, the passive (active) generators have a well-known interpretation: they conserve (do not conserve) photon number. Unlike active generators, which require nonlinear photon interactions, passive generators can be implemented with linear optical components: beam splitters and phase shifters. Now, despite the exact isomorphism between symplectic transformations on photon number states and spatial modes, they have quite distinct physical implications. To gain insight on how the metaplectic operator~(\ref{eq:metaplectic}) affects spatial modes, we resort to the Wigner representation in conjunction with the Stone-von Neumann theorem~(\ref{eq:StonevonNeumannTh}). A revealing example is the following. Consider a Laguerre-Gaussian mode~\cite{OAMbook} $\textrm{LG}_{\ell,p}$, where the indices $\ell=0,\pm 1,\pm 2,\ldots$ and $p=0, 1, 2,\ldots$ stand for the topological charge and the number of nonaxial radial nodes. Let $W_{\ell,p}({\boldsymbol \xi})$ be the associated Wigner function, which, in the general case, is non-Gaussian~\cite{Simon00a,Calvo05,Calvo06}. For each of the ten generators~(\ref{eq:passiveoperators}) and (\ref{eq:activeoperators}) one can easily obtain the corresponding symplectic matrix $S$ acting on the input $W_{\ell,p}({\boldsymbol \xi})$ to yield exact analytical expressions for the output Wigner functions (via its covariant property). The resulting position distribution $I_{\ell,p}^{(S)}({\bf q})$ is computed via the marginal $I_{\ell,p}^{(S)}({\bf q})=\int d^{2}{\bf p}W_{\ell,p}(S^{-1}{\boldsymbol \xi})$. Figure~\ref{fig:GeneratorAction} depicts $I_{\ell=1,p=0}^{(S)}({\bf q})$ under the action of each passive and active generator. It can be seen that they produce fundamentally different mode-mapping geometries. Passive generators describe rotations on the orbital Poincar\'{e} sphere~\cite{Calvo05,Calvo06,Padgett99}. They preserve the order $N\equiv\vert\ell\vert+2p$ of any of the modes lying on the sphere. Generator $\hat{\mathcal{L}}_{o}$ yields the mode-order, $\hat{\mathcal L}_{z}$ represents real spatial rotations on the transverse $x-y$ plane containing the modes and is proportional to the component of the OAM operator along the propagation direction~\cite{Calvo06}, with $\textrm{LG}_{\ell,p}$ being the eigenmodes. $\hat{\mathcal L}_{x}$ and $\hat{\mathcal L}_{y}$ represent simultaneous rotations in the four-dimensional phase-space: $\hat{\mathcal L}_{x}$ produces rotations in the $x-p_{x}$ and $y-p_{y}$ planes by equal and opposite amounts, whereas $\hat{\mathcal L}_{y}$ gives rise to rotations in the $x-p_{y}$ and $y-p_{x}$ planes by equal amounts. The eigenstates of $\hat{\mathcal L}_{x}$ are the Hermite-Gaussian $\textrm{HG}_{n_{x},n_{y}}$ modes, where $n_{x},n_{y}$ are nonnegative integer indices, and their mode-order is $N\equiv n_{x}+n_{y}$. Both Laguerre- and Hermite-Gaussian bases are unitarily related: $\textrm{LG}_{\ell,p}$ transforms into $\textrm{HG}_{n_{x},n_{y}}$ via $e^{-i(\pi/2)\hat{\mathcal L}_{y}}$. We note in passing that the interferometric scheme proposed in Ref.~\cite{Zambrini} to measure the OAM spectrum (i.e. the $\ell$ index-decomposition) of a light beam, which only involves $\hat{\mathcal{L}}_{z}$, could be generalized to determine the complete Hermite-Gaussian spectrum by replacing $\hat{\mathcal{L}}_{z}$ with $\hat{\mathcal{L}}_{o}$ and $\hat{\mathcal{L}}_{x}$~\cite{Calvo08}. In contrast with passive generators, the active ones scale ({\em squeeze}) the spatial modes and change the order $N$, giving rise to infinite mode-superpositions. Of these, only $\hat{\mathcal K}_{z}$ and $\hat{\mathcal M}_{y}$ suffice to describe, jointly with the set~(\ref{eq:passiveoperators}), the general metaplectic operator~(\ref{eq:metaplectic}) by recourse to the following passive-active-passive decomposition $\hat{U}=e^{-i{\boldsymbol\mu}\cdot\hat{\boldsymbol{\mathcal{L}}}}e^{-i(\nu_{y}\hat{\mathcal M}_{y}+\nu_{z}\hat{\mathcal K}_{z})}e^{-i{\boldsymbol \eta}\cdot\hat{\boldsymbol{\mathcal{L}}}}$, still requiring ten parameters. The symplectic matrices associated with both passive and active generators can be implemented with a small arrange of ($<10$) spherical and/or cylindrical lenses solely controlled by variations of the focal lengths and/or rotations along the system axis~\cite{Calvo08,Bastiaans,Rodrigo}; that is, with simple linear optical components. 
\par
\section{METAPLECTIC OPERATIONS ON SINGLE-PHOTON QUDITS}
\label{sec:SQGMO}
Having characterized the complete set of unitaries generated by passive~(\ref{eq:passiveoperators})  and active~(\ref{eq:activeoperators})  operators that can be performed on paraxial field modes, we
proceed with all possible actions of the metaplectic operator~(\ref{eq:metaplectic}) on arbitrary single-photon qu$d$it states. The most general (paraxial) single-photon pure state can be described as~\cite{Calvo06}
\begin{eqnarray}
\vert \psi\rangle = \sum_{\sigma,n_{x},n_{y}}\int_{0}^{\infty} d\omega\, C_{\sigma,n_{x},n_{y}}(\omega)\,\hat{b}_{\sigma,n_{x},n_{y}}^{\dagger}\left(\omega\right)\vert \textrm{vac} \rangle\; . 
\label{eq:onephotonparaxial}
\end{eqnarray}
Here, $\hat{b}_{\sigma,n_{x},n_{y}}^{\dagger}(\omega)$ denotes the bosonic creation operator of a Hermite-Gaussian mode, linear polarization $\sigma$, and frequency $\omega$. The commutation relations read as $[\hat{b}_{\sigma,n_{x},n_{y}}\left(\omega\right), \hat{b}_{\sigma',n_{x}',n_{y}'}^{\dagger}(\omega')]=\delta_{\sigma\sigma'}\delta_{n_{x}n_{x}'}\delta_{n_{y}n_{y}'}\delta(\omega-\omega')$. The normalized complex coefficients $C_{\sigma,n_{x},n_{y}}(\omega)$ can be interpreted as the probability amplitudes for finding a photon in the state $\hat{b}_{\sigma,n_{x},n_{y}}^{\dagger}\left(\omega\right)\vert \textrm{vac} \rangle=\vert \sigma\rangle\otimes\vert n_{x}, n_{y}\rangle\otimes\vert\omega\rangle$. Let us concentrate on the spatial part of Eq.~(\ref{eq:onephotonparaxial}) and assume that it consists of a {\em finite} superposition of $d$ orthogonal HG modes $\vert n_{x},n_{y}\rangle= (\hat{a}_{x}^{\dagger})^{n_{x}}(\hat{a}_{y}^{\dagger})^{n_{y}}\vert 0,0 \rangle$ ($\vert 0,0 \rangle$ is the fundamental Gaussian mode), in the normalized qu$d$it form
\begin{eqnarray}
\vert \psi \rangle =  \sum_{n_{x},n_{y}}c_{n_{x},n_{y}}\vert n_{x},n_{y}\rangle\, .
\label{eq:qudit}
\end{eqnarray}
Any qu$d$it requires, at least, $2d$ independent real parameters, albeit normalization and invariance of~(\ref{eq:qudit}) under a global phase reduces this number to $2(d-1)$. A necessary condition to fully manipulate a single-photon state~(\ref{eq:qudit}) is to arbitrarily modify the $d$ complex coefficients $c_{n_{x},n_{y}}$ (e.g. it should be possible to set all coefficients $c_{n_{x},n_{y}}$ equal to zero, except for one of them), leaving invariant the $d$-dimensional subspace $\mathcal{H}_{d}$ expanded by $\{\vert n_{x},n_{y}\rangle\}$. In other words, one must discard all transformations on~(\ref{eq:qudit}) giving rise to modes not belonging to $\mathcal{H}_{d}$. 
\par 
Let us analyze the most important restrictions imposed by unitary operations acting on~(\ref{eq:qudit}) and generated by the group of transformations $S\in\textrm{Sp}(4,\mathbf{R})$. First, notice that since the general metaplectic operator~(\ref{eq:metaplectic}) involves ten generators, $d$-dimensional superpositions with $d>6$ cannot be arbitrarily transformed within Sp$(4,\mathbf{R})$. This fact, of course, does not preclude the possibility to manipulate qu$d$its with $d\leq6$. A key observation is the recognition that finite dimensional representations of Sp(4,$\mathbf{R}$) are necessarily nonunitary, owing to the noncompactness of Sp(4,$\mathbf{R}$). That is, the noncompact part of Sp(4,$\mathbf{R}$), represented by the active generators~(\ref{eq:activeoperators}), is to be excluded from the set of symplectic transformations in order to maintain the subspace $\mathcal{H}_{d}$ finite. Otherwise, the qu$d$it~(\ref{eq:qudit}) would become an infinite superposition of all HG modes under the general action of~(\ref{eq:metaplectic}). More explicitly, let $\hat{\mathcal{P}}_{d}$ be a projector in $\mathcal{H}_{d}$, so that it fulfills $\hat{\mathcal{P}}_{d}\vert\psi\rangle=\vert\psi\rangle$. Notice that if the metaplectic operator~(\ref{eq:metaplectic}) must keep~(\ref{eq:qudit}) in $\mathcal{H}_{d}$, then $\hat{\mathcal{P}}_{d}\hat{U}\vert\psi\rangle=\hat{U}\vert\psi\rangle$, which implies that $\hat{U}^{-1}\hat{\mathcal{P}}_{d}\hat{U}=\hat{\mathcal{P}}_{d}$. This last condition should hold for any choice of the parameters ${\bf s}$ in~(\ref{eq:metaplectic}), and so it follows that the commutator $[{\bf s}\cdot\hat{\boldsymbol{\mathcal{J}}},\hat{\mathcal{P}}_{d}]=0$. However, this vanishing commutator is incompatible with the presence of the active generators~(\ref{eq:activeoperators}). One is therefore limited to the compact subgroup of Sp(4,$\mathbf{R}$), i.e. to the set of passive generators~(\ref{eq:passiveoperators}), to perform unitary operations on~(\ref{eq:qudit}) leaving the subspace $\mathcal{H}_{d}$ invariant. This means that the metaplectic operator~(\ref{eq:metaplectic}) reduces to the passive one $\hat{U}_{\mathcal{L}}=e^{-i(s_{o}\hat{\mathcal{L}}_{o}+s_{x}\hat{\mathcal{L}}_{x}+s_{y}\hat{\mathcal{L}}_{y}+s_{z}\hat{\mathcal{L}}_{z})}$, which, now, only contains the four free parameters $\{s_{o},s_{x},s_{y},s_{z}\}$. Consequently, one must confine to three-dimensional subspaces $\mathcal{H}_{d}$, with state~(\ref{eq:qudit}) being a qutrit. In principle, the intervening modes in~(\ref{eq:qudit}) can have different order $N$. However, since $\hat{U}_{\mathcal{L}}$ preserves $N$, this implies that it would then be impossible to attain the mode transformation $\vert n_{x}, n_{y} \rangle\to\vert n_{x}', n_{y}' \rangle $ when $n_{x}+n_{y}\neq n_{x}'+n_{y}'$. There is still the possibility that the three modes could have the same order. In this particular case, taking into account that $\hat{\mathcal{L}}_{o} \vert \psi \rangle = N/2 \vert \psi \rangle$ and $[\hat{\mathcal{L}}_{i},\hat{\mathcal{L}}_{o}] = 0$, one can express the action of $\hat{U}_{\mathcal{L}}$ as $\hat{U}_{\mathcal{L}}\vert \psi \rangle = e^{-is_{o}N/2}e^{-i(s_{x}\hat{\mathcal{L}}_{x}+s_{y}\hat{\mathcal{L}}_{y}+s_{z}\hat{\mathcal{L}}_{z})}\vert \psi \rangle$. Up to a global phase, there are only three free parameters $\{s_{x},s_{y},s_{z}\}$ to carry out the general transformations on the state~(\ref{eq:qudit}), which are insufficient even for qutrits (as they involve, at least, four free real parameters). We have thus proven the following result:
\par
{\em Proposition.-} It is impossible to arbitrarily modify the $d$-dimensional mode-superpositions of single-photon pure qu$d$it states~(\ref{eq:qudit}) for $d>2$, via unitary operations $\hat{U}(S)$ generated by symplectic transforms $S\in$ Sp(4,$\mathbf{R}$).
\par
Several comments are in order. Our proposition is expected to also hold for mixed states. Arbitrary operations on qubits ($d=2$) are not prohibited within the subgroup U(2) of Sp(4,$\mathbf{R}$), as it should be~\cite{Calvo06}. Operators $\hat{U}_{\mathcal{L}}$ acting on higher-than-two mode superpositions~(\ref{eq:qudit}) restrict the possible values of coefficients $c_{n_{x},n_{y}}$; the higher the value of $d$ the larger the number of constraints on $c_{n_{x},n_{y}}$. In fact, it is easy to show that, within Sp(4,$\mathbf{R}$), the most general {\em finite} superpositions~(\ref{eq:qudit}) reduce to the well-known Dicke or spin coherent states~\cite{Arecchi} (which depend only on two real parameters, leaving aside global phases). Although we have identified spherical and cylindrical lenses as the main optical elements of Sp(4,$\mathbf{R}$), the above proposition also affects phase holograms belonging to the metaplectic representation of Sp(4,$\mathbf{R}$). Not all unitary and paraxial transformations are in such representation. It is an open question to determine the entire class of unitary operators outside the metaplectic representation of Sp(4,$\mathbf{R}$) that leave invariant the bases of paraxial modes; but it would definitely fall into the category of non-Gaussian operations. In the next Section we partially clarify this question; we find examples of non-Gaussian operations which preserve paraxiallity and allow us to overcome the limit established by the preceding proposition. Moreover, since Laguerre- and Hermite-Gaussian bases are unitarily connected,  our proposition also establishes the impossibility to achieve arbitrary qu$d$it gates on multi-dimensional superpositions of modes bearing OAM. Given that the concept of OAM of light is only strictly meaningful within the paraxial approximation~\cite{OAMbook,Calvo06}, encoding photons in non-paraxial modes would generally couple polarization and OAM making most QC tasks in such scenarios extremely difficult. 
\par
\section{NON-GAUSSIAN OPERATIONS}
\label{sec:NGGates}
In contrast with Gaussian operations, non-Gaussian operations remain to be fully explored. It has been recognized that non-Gaussian operations could represent an advantage to perform some quantum information tasks. Non-Gaussian operations on continuous variables allow the access to results beyond no-go statements concerning Gaussian operations. For instance, quantum speed up is impossible for harmonic oscillators by Gaussian operations with Gaussian inputs~\cite{Bartlett}. Distillation of Gaussian bipartite entanglement is also impossible by performing only Gaussian local operations and classical communication based on homodyne detection and requires non-Gaussian operations~\cite{Distillation}. It has been experimentally demonstrated that entanglement between Gaussian entangled states can be increased by conditional subtraction of single photons from the Gaussian beams~\cite{Ourjoumtsev}. For universal quantum computation with continuous variable cluster states~\cite{Menicucci} it is necessary, at least, one non-Gaussian projective measurement. 
\par
In the present scenario, our preceding proposition imposes a restrictive limit for Gaussian operations on arbitrary superpositions of spatial mode states. However, in this section, we report a new class of non-Gaussian operations which enable us to fully manipulate superpositions of three-level states (qutrits), beyond the restrictions imposed by our above no-go proposition. Furthermore, this class of operations forms a complete set of single qutrit gates fulfilling a SU(3) algebra.
\par
\begin{figure}
\begin{center}
\hspace*{-0.0cm}
\includegraphics[width=60mm]{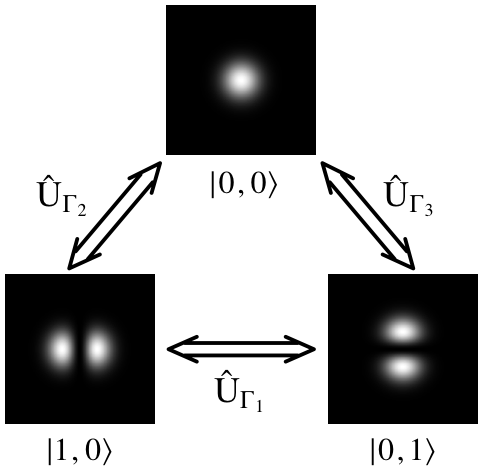}
\end{center}
\vspace*{-0.3cm}
\caption{\small Mode conversion via the unitary operations $\hat{U}_{\Gamma}$ generated by the SU(3) group~(\ref{eq:non-Gaussian_oper}).}
\label{fig:NonGaussianConversion}
\end{figure}

Consider the eight following generators acting on the Hilbert space of Hermite-Gaussian modes:
\begin{eqnarray}
\hat{\mathcal{T}}_{1}&=& \frac{1}{2} \left(\hat{a}_{x}^{\dagger}\hat{a}_{y}+\hat{a}_{y}^{\dagger}\hat{a}_{x}\right) , \nonumber \\
\hat{\mathcal{T}}_{2}&=&  - \frac{i}{2} \left(\hat{a}_{x}^{\dagger}\hat{a}_{y}-\hat{a}_{y}^{\dagger}\hat{a}_{x}\right) , \nonumber \\
\hat{\mathcal{T}}_{3}&=& \frac{1}{2} \left(\hat{a}_{x}^{\dagger}\hat{a}_{x}-\hat{a}_{y}^{\dagger}\hat{a}_{y}\right) , \nonumber \\
\hat{\mathcal{T}}_{4}&=& \frac{1}{2} \left(\hat{a}_{x}^{\dagger}+\hat{a}_{x}-\hat{a}_{x}^{\dagger}\hat{a}_{x}^{\dagger}\hat{a}_{x}-\hat{a}_{x}^{\dagger}\hat{a}_{x}\hat{a}_{x} \right.\nonumber \\
&-& \left. \hat{a}_{x}^{\dagger}\hat{a}_{y}^{\dagger}\hat{a}_{y} - \hat{a}_{x}\hat{a}_{y}^{\dagger}\hat{a}_{y} \right) , \nonumber \\
\hat{\mathcal{T}}_{5}&=& -\frac{i}{2} \left(\hat{a}_{x}^{\dagger}-\hat{a}_{x}-\hat{a}_{x}^{\dagger}\hat{a}_{x}^{\dagger}\hat{a}_{x}+\hat{a}_{x}^{\dagger}\hat{a}_{x}\hat{a}_{x} 
\right.\nonumber \\
&-& \left. \hat{a}_{x}^{\dagger}\hat{a}_{y}^{\dagger}\hat{a}_{y} + \hat{a}_{x}\hat{a}_{y}^{\dagger}\hat{a}_{y} \right) , \nonumber \\
\hat{\mathcal{T}}_{6}&=&  \frac{1}{2} \left(\hat{a}_{y}^{\dagger}+\hat{a}_{y}-\hat{a}_{y}^{\dagger}\hat{a}_{y}^{\dagger}\hat{a}_{y}-\hat{a}_{y}^{\dagger}\hat{a}_{y}\hat{a}_{y} 
\right.\nonumber \\
&-& \left. \hat{a}_{y}^{\dagger}\hat{a}_{x}^{\dagger}\hat{a}_{x} - \hat{a}_{y}\hat{a}_{x}^{\dagger}\hat{a}_{x} \right) , \nonumber \\
\hat{\mathcal{T}}_{7}&=& -\frac{i}{2} \left(\hat{a}_{y}^{\dagger}-\hat{a}_{y}-\hat{a}_{y}^{\dagger}\hat{a}_{y}^{\dagger}\hat{a}_{y}+\hat{a}_{y}^{\dagger}\hat{a}_{y}\hat{a}_{y} 
\right.\nonumber \\
&-& \left. \hat{a}_{y}^{\dagger}\hat{a}_{x}^{\dagger}\hat{a}_{x} + \hat{a}_{y}\hat{a}_{x}^{\dagger}\hat{a}_{x} \right) , \nonumber \\
\hat{\mathcal{T}}_{8}&=& \frac{1}{2\sqrt{3}}\left[-2+3\left(\hat{a}_{x}^{\dagger}\hat{a}_{x}+\hat{a}_{y}^{\dagger}\hat{a}_{y}\right)\right] .
\label{eq:non-Gaussian_oper}
\end{eqnarray}
Quite remarkably, these generators, within the subspace expanded by the Hermite Gaussian modes $\mathcal{H}_{\mathcal{T}}=\{\vert 0, 0 \rangle,\vert 1, 0 \rangle,\vert 0, 1 \rangle\}$, obey the SU(3) algebra $[\hat{\mathcal{T}}_{a},\hat{\mathcal{T}}_{b}]= if_{abc}\hat{\mathcal{T}}_{c}$ ($a,b,c=1,2,\ldots,8$), where the only nonvanishing (up to permutations) structure constants $f_{abc}$ are given by $f_{123}=1$, $f_{147}=f_{165}=f_{246}=f_{257}=f_{345}=f_{376}=1/2$, and $f_{458}=f_{678}=\sqrt{3}/2$. Unlike in the metaplectic representation of Sp(4,$\mathbf{R}$), there are now four generators in~(\ref{eq:non-Gaussian_oper}) involving cubic terms in $\hat{a}_{x}, \hat{a}_{x}^{\dagger}, \hat{a}_{y}, \hat{a}_{y}^{\dagger}$. Their structure resembles that of fully-quantum interaction Hamiltonians found in nonlinear parametric processes. Notice that the triad of generators $\Gamma_{1}\equiv\{\hat{\mathcal{T}}_{1},\hat{\mathcal{T}}_{2},\hat{\mathcal{T}}_{3}\}$ corresponds to the SU(2) group in Eq.~(\ref{eq:passiveoperators}). The remaining two SU(2) groups are formed by the triads: $\Gamma_{2}\equiv\{\hat{\mathcal{T}}_{4},\hat{\mathcal{T}}_{5}, (\hat{\mathcal{T}}_{3}+\sqrt{3}\hat{\mathcal{T}}_{8})/2\}$, and $\Gamma_{3}\equiv\{\hat{\mathcal{T}}_{6},\hat{\mathcal{T}}_{7}, (-\hat{\mathcal{T}}_{3}+\sqrt{3}\hat{\mathcal{T}}_{8})/2\}$. Unitary operators $\hat{U}_{\Gamma_{1}}$ generated by the first triad give rise to superpositions between the two modes $\vert 1, 0 \rangle$ and $\vert 0, 1 \rangle$, leaving invariant the fundamental Gaussian mode $\vert 0, 0 \rangle$. Unitaries $\hat{U}_{\Gamma_{2}}$ and $\hat{U}_{\Gamma_{3}}$, generated by the second and third triad, produce superpositions between the two modes $\vert 0, 0 \rangle$ and $\vert 1, 0 \rangle$ (leaving invariant $\vert 0, 1 \rangle$),  or the modes $\vert 0, 0 \rangle$ and $\vert 0, 1 \rangle$ (leaving invariant $\vert 1, 0\rangle$), respectively. In contrast with $\hat{U}_{\Gamma_{1}}$, the action of $\hat{U}_{\Gamma_{2}}$ and $\hat{U}_{\Gamma_{3}}$ on $\mathcal{H}_{\mathcal{T}}$ gives rise to a new feature: non-conservation of the mode order. However, this non-conservation is fundamentally different to the one encountered in the noncompact representation of Sp(4,$\mathbf{R}$), since it preserves the subspace $\mathcal{H}_{\mathcal{T}}$. Figure~\ref{fig:NonGaussianConversion} summarizes the action of the three unitaries $\hat{U}_{\Gamma}$ on the subspace $\mathcal{H}_{\mathcal{T}}$.  It is worth mentioning that all these non-Gaussian operations can be implemented with passive optical elements having higher-than-first-order aberrations (non-quadratic refractive surfaces)~\cite{ABERRATIONS}. An open and interesting problem would be the extension of the Stone-von Neumann theorem~(\ref{eq:StonevonNeumannTh}) to the case of our qubic generators~(\ref{eq:non-Gaussian_oper}). This would enable one to find the explicit form of the symplectic transform and thus the construction of the associated optical system.
\par
Complete manipulation of the qutrit $\vert\psi\rangle= c_{0,0}\vert 0, 0\rangle+c_{1,0}\vert 1, 0\rangle+c_{0,1}\vert 0, 1\rangle$ is now possible using the SU(3) group~(\ref{eq:non-Gaussian_oper}), although to produce a general qutrit (up to a global phase) it suffices to perform the following sequence of operations: starting, for example, with an input fundamental Gaussian mode $\vert 0, 0 \rangle$ subjected to the unitary operator $\hat{U}_{\Gamma_{2}}$, one can obtain
\begin{eqnarray}\label{eq:qubit1}
\hat{U}_{\Gamma_{2}}\vert 0, 0 \rangle = \cos{\frac{\theta}{2}} \vert 0, 0 \rangle + e^{i \varphi } \sin{\frac{\theta}{2}} \vert 1, 0\rangle \,.
\end{eqnarray}
Then, taking into account the closed SU(2) algebra obeyed by $\hat{U}_{\Gamma_{1}}$, it follows that
\begin{eqnarray}\label{eq:qubit2}
 \hat{U}_{\Gamma_{1}}\vert 1, 0 \rangle = \cos{\frac{\theta'}{2}} \vert 1, 0 \rangle + e^{i \varphi' } \sin{\frac{\theta'}{2}} \vert 0, 1\rangle \,.
\end{eqnarray}
With this specific operation structure, using (\ref{eq:qubit1}) and (\ref{eq:qubit2}), we can construct a general normalized qutrit state (up to a global phase):
\begin{eqnarray} 
\hat{U}_{\Gamma_{1}}\hat{U}_{\Gamma_{2}}\vert 0, 0 \rangle &=& \cos{\frac{\theta}{2}} \vert 0, 0 \rangle + e^{i \varphi } \sin{\frac{\theta}{2}} \cos{\frac{\theta'}{2}} \vert 1, 0 \rangle \nonumber \\ &+& e^{i( \varphi + \varphi')} \sin{\frac{\theta}{2}} \sin{\frac{\theta'}{2}} \vert 0, 1\rangle \, . \label{eq:qutrit}
\end{eqnarray}
Since the four parameters $\theta,\theta',\varphi,\varphi'$ can be varied independently during the process, not all generators in the set~(\ref{eq:non-Gaussian_oper}) are actually needed to produce any qutrit encoded only in paraxial spatial modes. Notice that our results can also be extended to other physical scenarios, due to the isomorphism of the formalism, although in our case an additional motivation is provided by the simplicity of their experimental implementation. 
\par
\begin{figure}
\begin{center}
\hspace*{-0.0cm}
\includegraphics[width=80mm]{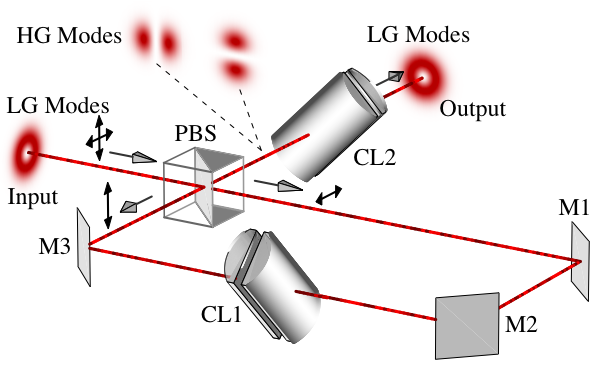}
\end{center}
\vspace*{-0.4cm}
\caption{\small (Color online) Scheme of a single-photon polarization-CNOT gate. According to the polarization of the input Laguerre-Gaussian mode the photon experiences a sign change ($\ell\to-\ell$, where $\vert\ell\vert=1$) in its OAM.}
\label{fig:SCNOT}
\end{figure}

\section{DISCUSSION AND CONCLUSIONS}
\label{sec:CONCLU}
In spite of the stringent limits raised by our results of Section~\ref{sec:SQGMO}, we have clearly shown how non-Gaussian operations can circumvent most of those difficulties. It is also worth emphasizing that there exist other alternative approaches, exploiting the spatial encoding of light, to fully manipulate higher-than-two-dimensional Hilbert spaces for various quantum information tasks. These approaches rely on the use of several degrees of freedom of light, albeit they cannot attain very large subspace-dimensionalities. To illustrate, consider that we perform transformation~(\ref{eq:qubit2}) and, analogously to transformation (\ref{eq:qubit1}), we wish to produce a general (up to a global phase) single-photon qutrit state. Instead of (\ref{eq:qubit1}), we can use another degree of freedom of the same photon (e.g. polarization). To do so, we need both a complete set of single-photon qubit gates in each of the degrees of freedom and a conditional gate between the two involved degrees of freedom. For instance,  an efficient linear single-photon CNOT gate in which photon polarization acts as the control-qubit on the other photon degree of freedom, OAM, which plays the role of the target-qubit, is feasible with current technology.  There are several possible routes, such as with space-variant optical axis phase plates made of nematic liquid-crystals~\cite{Marrucci,Calvo07b}, or using a Mach-Zender configuration~\cite{Deng}. A conceptually simple scheme is depicted in Fig.~\ref{fig:SCNOT}. This gate includes one polarizing beam splitter (PBS) and two pairs of cylindrical lenses (CL) whose bases subtend a 45$^{\circ}$ angle with the plane of the interferometer. This interferometer resembles previous Sagnac interferometers used for measuring the spatial Wigner function~\cite{Walmsley03} and for other single-photon quantum gate demonstrations employing polarization and continuous variables~\cite{Fiorentino}. There, the inner arm of the interferometer contained a Dove prism. Here, the presence of cylindrical lenses constitutes the key feature to exploit photon OAM. According to the input photon polarization state (horizontal or vertical), the input photon views the cylindrical lenses CL1 with a different orientation and experiences a mode transformation ($LG\rightarrow HG$) depending on the particular value of $\ell$. After exiting through the PBS, the second cylindrical lenses CL2 yields the mode transformation $HG\rightarrow LG$, which completes the action of the polarization-OAM-CNOT single-photon gate.
\par
An even more fascinating scenario is the transfer of photons carrying OAM onto Bose-Einstein condensates~\cite{BEC} and their storage in electromagnetically induced transparency media~\cite{EIT}. In this respect, it would be very interesting to explore the possibility of mapping the correlations of photons entangled in OAM~\cite{Mair,Vaziri03,Molina04,Langford,Kumar,Aiello05,Calvo07} on quantum holograms, allowing for the reconstruction of nonclassical states of light from a matter-based quantum memory.
\par
In conclusion, we have shown that if single-photon pure qu$d$it states are prepared in a $d$-dimensional superposition of spatial modes, it is impossible to arbitrarily change such mode-superpositions for $d>2$, by solely resorting to unitary operations generated by symplectic transforms of the group Sp(4,$\mathbf{R}$). Our results provide a complete characterization of linear canonical transformations on transverse optical modes and pose a considerable challenge on quantum communication protocols exploiting multidimensional spatial encoding: one cannot have a full access and control of large but finite-dimensional Hilbert spaces expanded by these modes. Implementation of a new class of paraxial non-Gaussian transformations to fully encode any arbitrary single-photon qu$d$it state is required. We have provided an explicit construction of this new class of operations. Moreover, using the spatial encoding in combination with other degrees of freedom, one can overcome this problem though at the price of scalability. In this case, a conditional gate between the involved degrees of freedom is needed. 
\par
We thank useful discussions with G. Giedke and acknowledge financial support from the Spanish Ministerio de Educaci\'{o}n y Ciencia through the Juan de la Cierva Grant Program and Projects FIS2005-01369 and Consolider Ingenio 2010 QIOT CSD2006-00019.

\end{document}